\documentclass{PoS}
\usepackage{wrapfig}
\usepackage{amsmath}
\usepackage{epsfig}

\title{Diffractive Dijet Production with a Leading Proton in ep Collisions at HERA}

\ShortTitle{Diffractive Dijet Production with a Leading Proton at HERA}

\author{\speaker{Stefan Schmitt}\thanks{On behalf of the H1 collaboration}\\
        DESY, Notekestr.~85, 22607 Hamburg, Germany\\
        E-mail: \email{sschmitt@mail.desy.de}}

\abstract{The production of dijets with a tagged forward proton is
  measured at HERA. The data were recorded with the 
  H1 detector at DESY in the years 2006-2007. Events with a leading
  proton are detected using the very forward proton spectrometer of the H1
  detector. Two jets are selected with transverse momenta transverse
  momenta in the hadronic-centre-of-mass larger than
  $4$ and $5.5\,\text{GeV}$, respectively. The analysis is performed
  both in the regime of deep-inelastic scattering (DIS), with momentum
  transfer $Q^2>4\,\text{GeV}^2$ and for photoproduction ($\gamma p$), with
  $Q^2<2\,\text{GeV}^2$. Cross sections are measured
  single-differentially in various kinematic quantities. For DIS, the data are 
  found to be in good agreement with NLO QCD calculations based on
  diffractive parton densities determined from inclusive diffractive
  cross section measurements. For $\gamma p$, the cross sections are
  found to be overestimated by approximately a factor of two.}

\FullConference{XXIII International Workshop on Deep-Inelastic Scattering,\\
		27 April - May 1 2015\\
		Dallas, Texas}

\begin{document}
\section{Introduction}

At HERA, reactions of electrons or positrons and protons, $ep\to eX$,
are probed at centre-of-mass energies of $320\,\text{GeV}$. Two
kinematic regimes are distinguished, depending on the  negative
momentum transfer squared $Q^2$ from the ingoing
to the outgoing electron\footnote{Throughout this paper, the term
  electron or the variable $e$ is used to denote both electrons and
  positrons, unless otherwise stated.}.
At high $Q^2>4\,\text{GeV}^2$, the process is referred to as
deep-inelastic scattering (DIS), whereas at low $Q^2<2\,\text{GeV}^2$
the process is called photoproduction ($\gamma p$).
In the analysis presented here \cite{Andreev:2015cwa}, dijet
production in both diffractive $\gamma p$ and diffractive DIS is
studied.
The reaction  may be written as $ep\to eXp$, where $X$ is a hadronic
system which contains two jets. The outgoing proton
is detected in the H1 Very Forward Proton Spectrometer (VFPS)
\cite{Astvatsatourov:2014dna}.

\begin{wrapfigure}{r}{0.4\textwidth}
\begin{center}
\epsfig{file=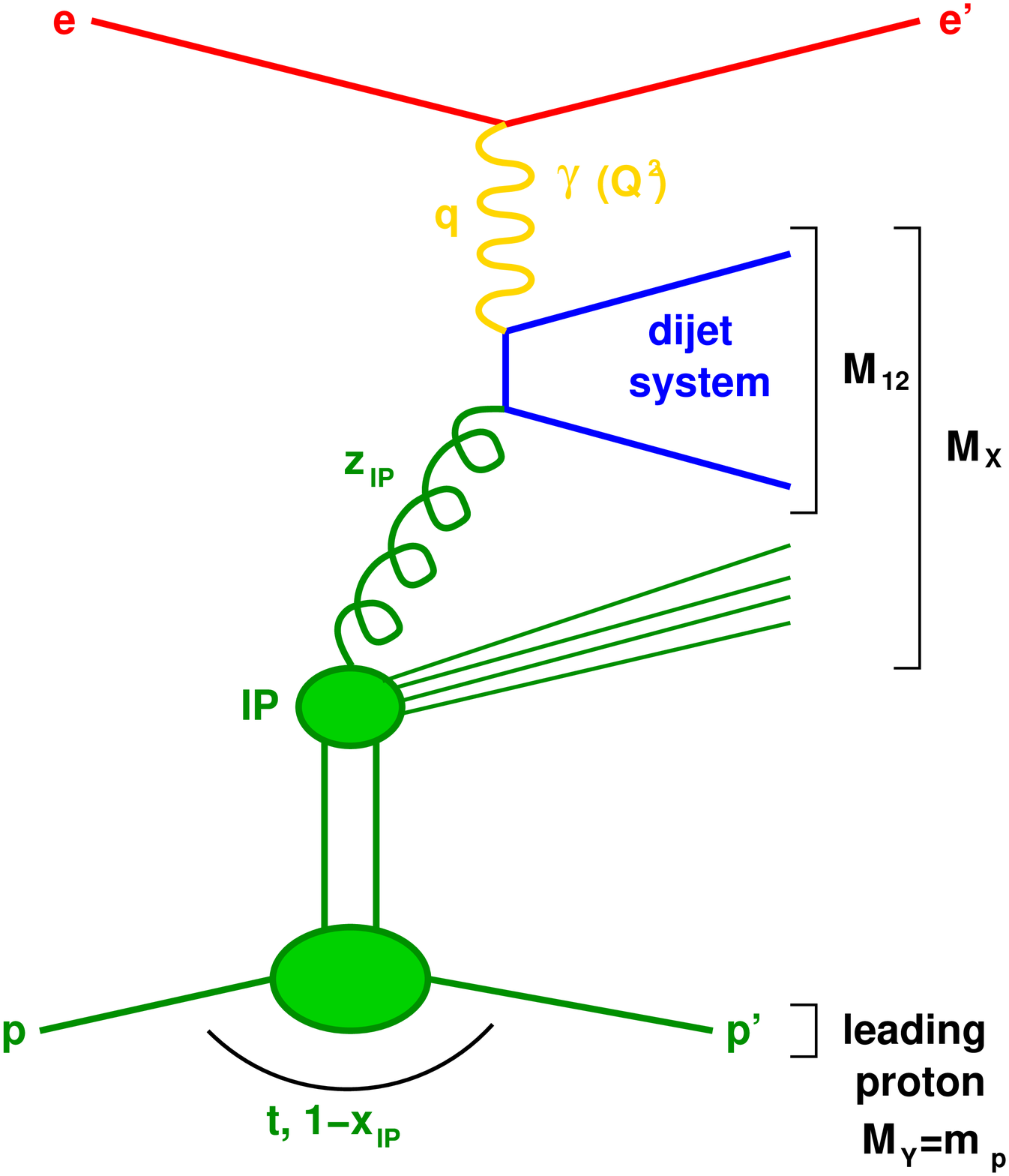,width=0.4\textwidth}
\caption{\label{fig:vfpsdijet}Diffractive dijet production with a
  leading proton at HERA.}
\end{center}
\end{wrapfigure}
QCD calculations for this process are based on diffractive parton
densities (DPDFs).
A corresponding factorisation theorem is known for the case of
diffractive DIS \cite{Collins:1997sr} but not for diffractive $\gamma p$.
The DPDFs describe the probability to find a parton with longitudinal
momentum fraction $z_{I\-\-P}$ in the proton, given that there is a
diffractive signature, characterised by a longitudinally momentum
fraction $1-x_{I\-\-P}$, momentum transfer $t$ of the outgoing proton
and a hard scale $\mu$.
They are folded with hard matrix elements to describe jet production at
next-to-leading order in the strong coupling.
A hard scale $\mu^2=\langle E_T^{\star jet}\rangle^2+Q^2$ is provided by the jet transverse momentum $\langle E_T^{\star jet}\rangle$ and in the case of DIS by the momentum transfer $Q^2$.

Experimentally, the DPDFs are determined from inclusive diffractive
DIS cross section measurements, where no
requirements on the hadronic final state $X$ are made.
As suggested by the scheme shown in figure \ref{fig:vfpsdijet}, the
DPDFs are determined with the ad-hoc assumption that they factorise into a
probability to find a colourless object $I\-\-P$ in the
proton and parton density functions, ascribed to the structure of
$I\-\-P$. The probability or ``flux factor'' is taken to depend on
$x_{I\-\-P}$ and $t$ only, 
whereas the parton density functions of $I\-\-P$ only depend on the
variable $z_{I\-\-P}$ and the hard scale $\mu$.

For the present analysis,
the H1 2006 DPDF fit B \cite{Aktas:2006hy} is used to predict cross sections.
The present measurements focuses on testing the factorisation
assumptions both in the DIS and in the $\gamma p$ kinematic domain.
Factorisation in diffractive DIS has been validated by experiment 
\cite{Chekanov:2002qm,Aktas:2006up,Aktas:2007bv,Chekanov:2007aa,Aaron:2011mp,Andreev:2014yra}.
In hadron-hadron collisions clear evidence for a suppression of the
cross section as compared to the expectation by about one order of
magnitude is observed
\cite{Affolder:2000vb,Chatrchyan:2012vc}.
For diffractive $\gamma p$, dijet production has been measured at HERA
\cite{Aktas:2007hn,Chekanov:2007rh,Aaron:2010su}.
A suppression of about a factor of two is observed in the two independent
analyses by the H1 experiment, whereas in the ZEUS analysis
no suppression is seen \cite{Chekanov:2007rh}.

\section{Data Analysis}

\begin{wraptable}{r}{0.5\textwidth}
{\footnotesize
\begin{center}
\begin{tabular}{c|cc}
\hline
 & Photoproduction & DIS \\
\hline
Event & $Q^2<2\,\text{GeV}^2$ & $4<Q^2<100\,\text{GeV}^2$ \\ 
kinematics    & \multicolumn{2}{c}{$0.2<y<0.7$} \\
\hline
Leading & \multicolumn{2}{c}{$0.01<x_{I\-\-P}<0.024$} \\
proton & \multicolumn{2}{c}{$\vert t\vert<0.6\,\text{GeV}^2$} \\
& & $z_{I\-\-P}<0.8$ \\
\hline
Dijets & \multicolumn{2}{c}{$E_{T}^{\star jet1}>5.5\,\text{GeV}$} \\
 & \multicolumn{2}{c}{$E_{T}^{\star jet2}>4\,\text{GeV}$} \\
 & \multicolumn{2}{c}{$-1<\eta^{\text{jet1,2}}<2.5$} \\
\hline
\end{tabular}
\end{center}
}
\caption{\label{tab:selection}Analysis phase space.}
\end{wraptable}
%
The analysis \cite{Andreev:2015cwa} is based on events recorded with
the H1 detector \cite{Abt:1996hi} in the 
years 2006 and 2007, corresponding to an integrated luminosity of
$30\,\text{pb}^{-1}$ for $\gamma p$ and $50\,\text{pb}^{-1}$ for DIS.
DIS events are identified by the presence of an electron in
the H1 rear calorimeter (SpaCal) \cite{Appuhn:1996na}, whereas $\gamma
p$ events are identified by the absence of an electron in the SpaCal
or the liquid-argon calorimeter.
These requirements limit the accessible kinematic range to
$4<Q^2<80\,\text{GeV}^2$ for DIS and to $Q^2<2\,\text{GeV}^2$ for
$\gamma p$.
The inelasticity $y$ is also restricted, in order to ensure that
both the scattered electron and the hadronic final state are well measured. 
Diffractive events with a leading proton are selected by
requiring a tagged proton in the H1 VFPS detector
\cite{Astvatsatourov:2014dna}.
\begin{figure}[t]
\begin{center}
\epsfig{file=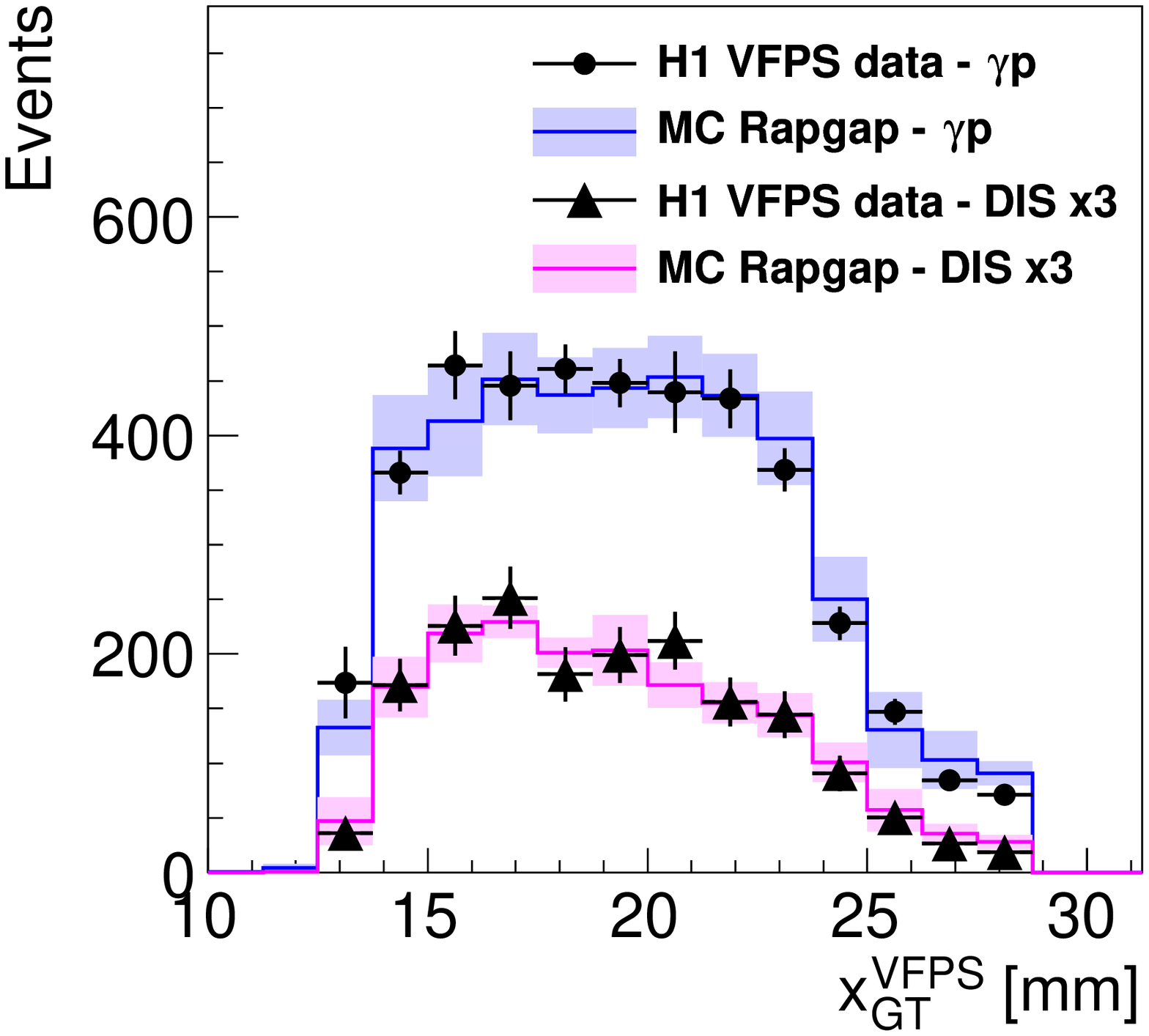,width=0.25\textwidth}%
\epsfig{file=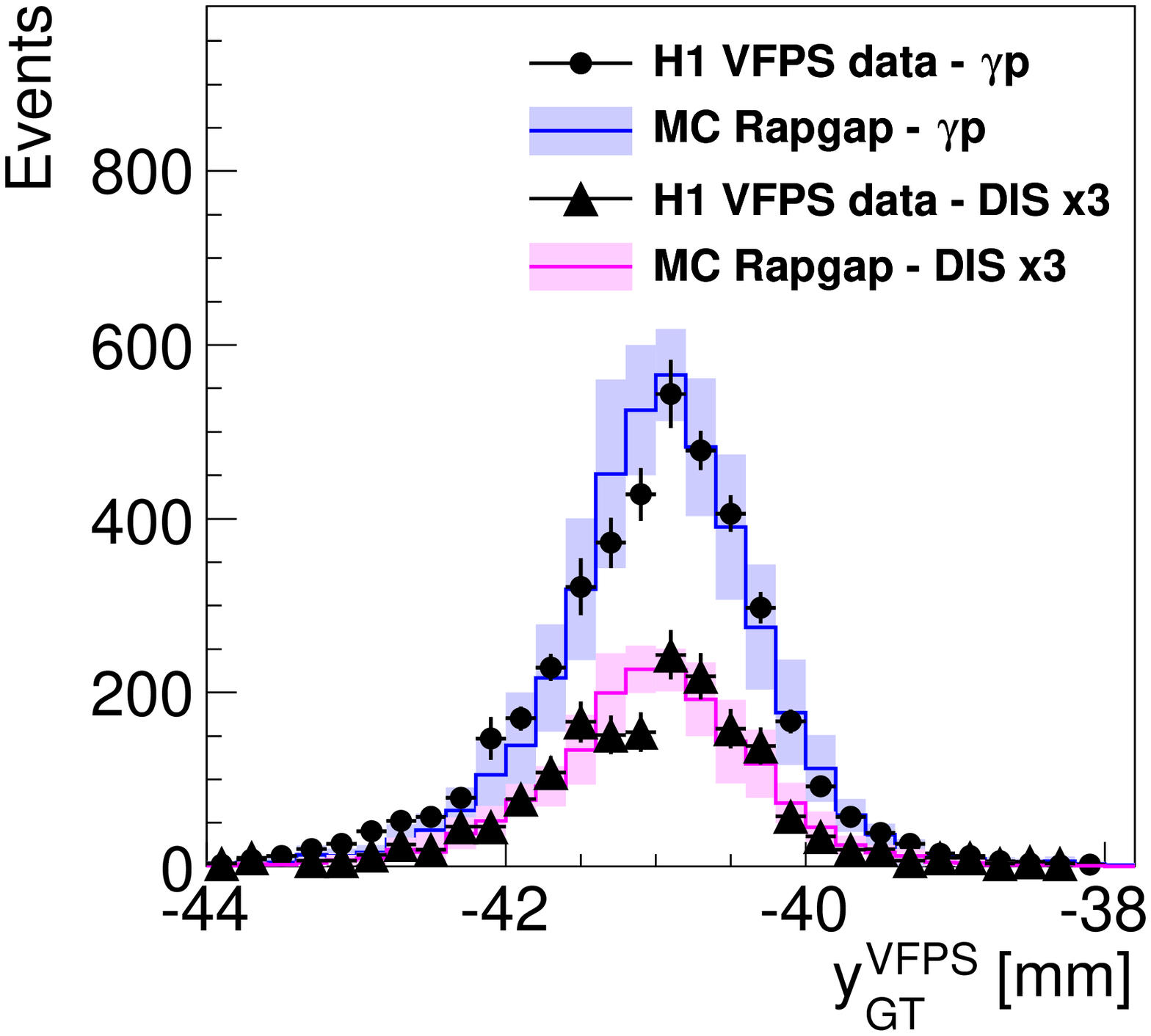,width=0.25\textwidth}%
\epsfig{file=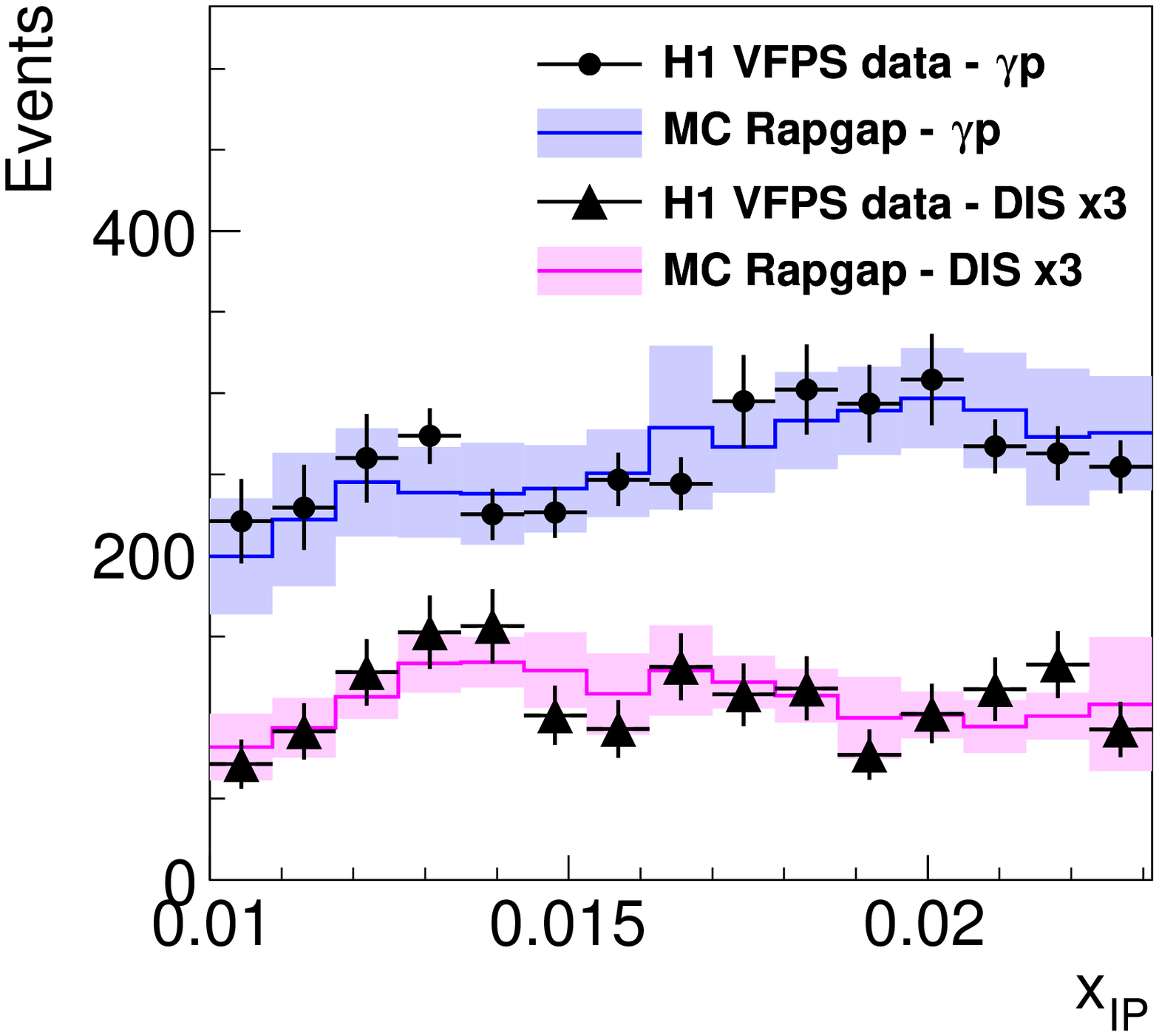,width=0.25\textwidth}
\epsfig{file=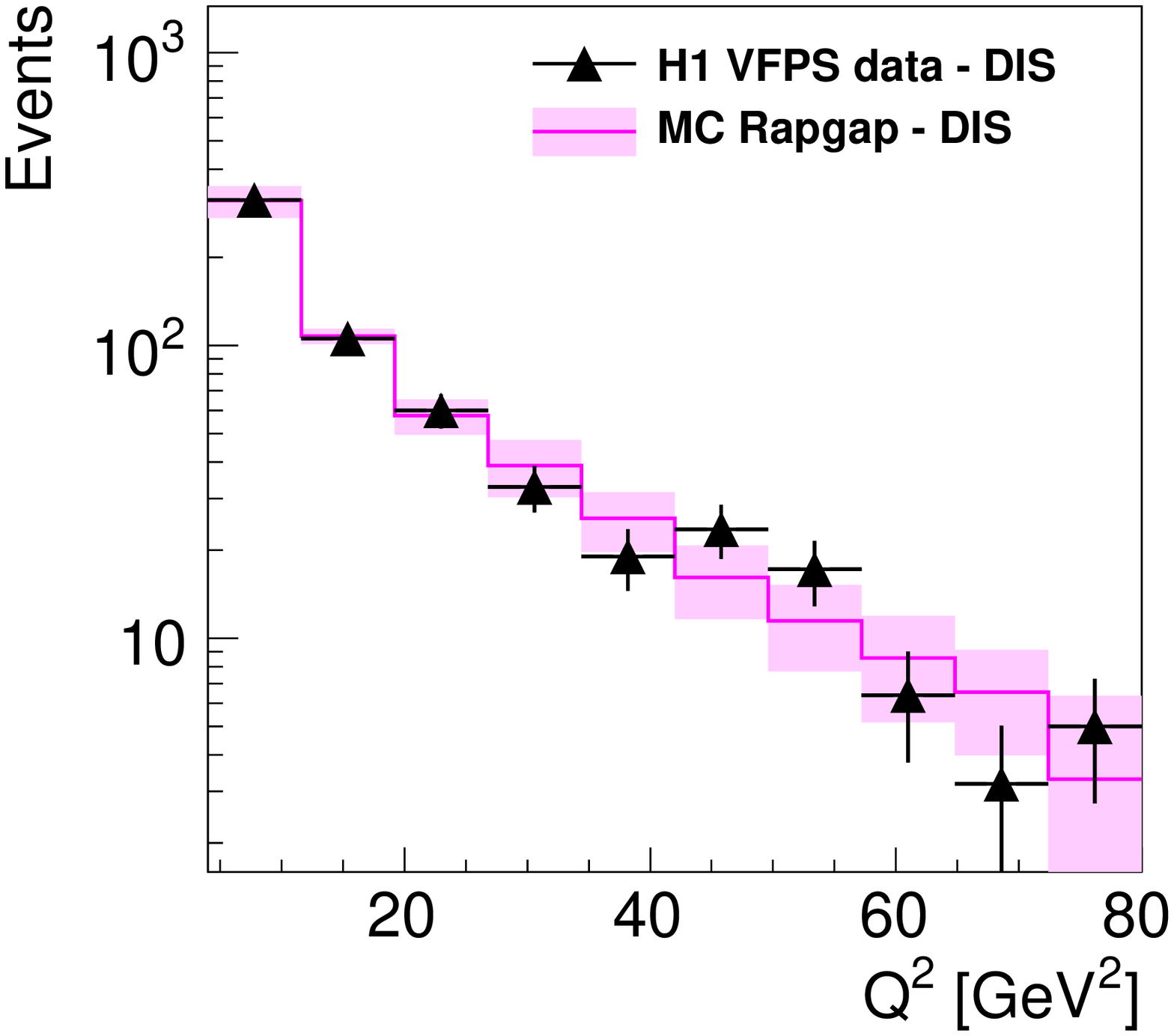,width=0.25\textwidth}%
\epsfig{file=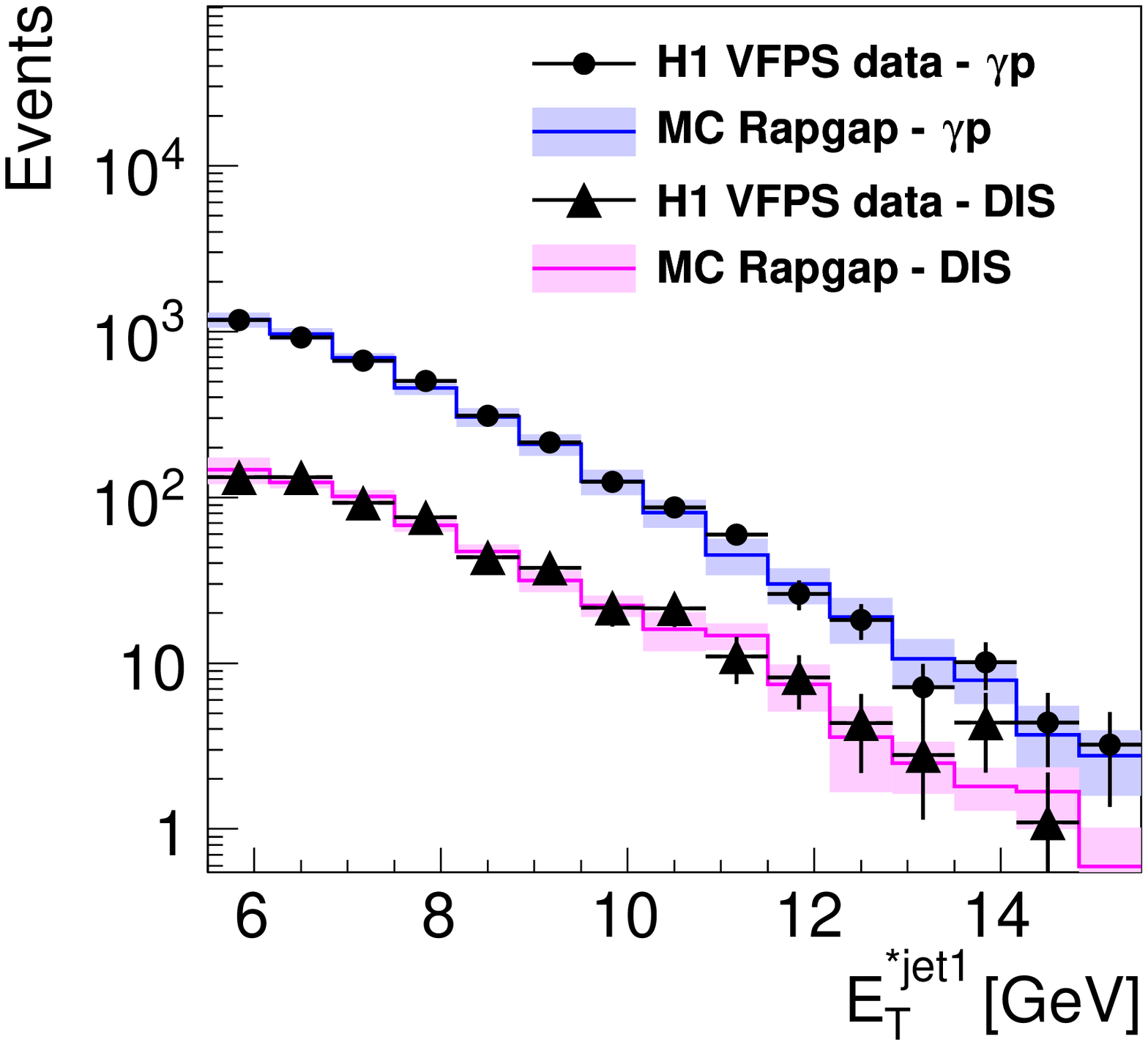,width=0.25\textwidth}%
\epsfig{file=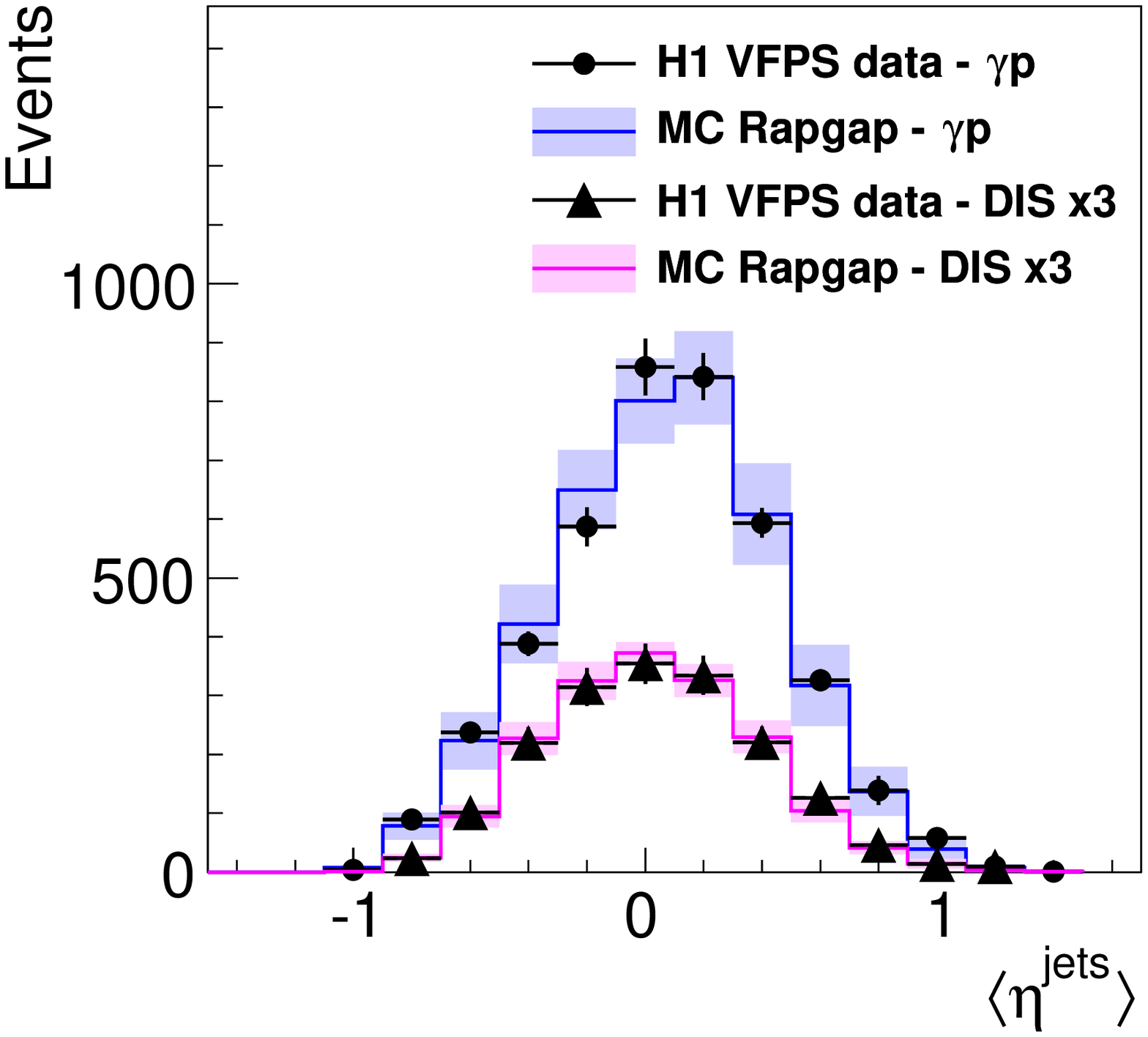,width=0.25\textwidth}
\caption{\label{fig:control}Control distributions of selected
  reconstructed quantities.}
\end{center}
\end{figure}
The VFPS has a high geometrical
acceptance (full coverage in azimuth),
but restricts the accessible range in
the forward proton longitudinal momentum $x_{I\-\-P}$ and in the
momentum transfer at the proton vertex $\vert t\vert$.
All selection criteria are summarised in table \ref{tab:selection}.
The hadronic final state $X$ is reconstructed from the detected tracks and 
calorimeter clusters, using an energy-flow algorithm.
The four-vectors of the hadronic objects are boosted to the $\gamma^{\,\star}\-p$ 
rest-frame and jets are reconstructed using the inclusive $k_T$ jet
algorithm \cite{Catani:1992zp} with $P_T$ recombination scheme and
distance parameter $R=1$.
Dijet events are accepted if there are at least two jets with
transverse momenta of the leading (sub-leading) jet fulfilling the conditions
$E_{T}^{\star jet1}>5.5\,\text{GeV}$ ($E_{T}^{\star
  jet2}>4\,\text{GeV}$). The jet pseudorapidity is 
restricted in the laboratory frame to $-1<\eta^{jet1,2}<2.5$, in
order to ensure that the jets are well contained\footnote{%
The $z$ axis is pointing along the proton
  flight direction. Polar angles $\theta$ are measured with respect to
  the $z$ axis. The pseudorapidity is defined as $\eta=-\ln\tan (\theta/2)$.}.
The events are selected in an extended analysis phase space, in
order to ensure that migrations near the phase space boundaries are be
well controlled.
Only after correcting for detector effects, the phase space
is restricted to the boundaries described in the text.
For unfolding from detector objects to the particle level, a
regularised matrix unfolding technique is applied.
The response matrix is constructed using the RAPGAP event
generator interfaced to the GEANT-based simulation of the H1 detector.
Control distributions of reconstructed variables at detector level are
shown in figure \ref{fig:control}.
The variables shown are: the local VFPS coordinates $X^{GT}_{VFPS}$
and $Y^{GT}_{VFPS}$, the proton fractional momentum loss $x_{I\-\-P}$ measured with
the VFPS, the momentum transfer $Q^2$
(DIS events only), the leading jet transverse momentum $E_{T}^{\star
  jet1}$ and the average jet pseudorapidity $\langle\eta^{jet}\rangle$.
All reconstructed variables are
well described by the RAPGAP simulation.

\section{Cross section measurements}

\begin{wrapfigure}{r}{0.3\textwidth}
\begin{center}
\epsfig{file=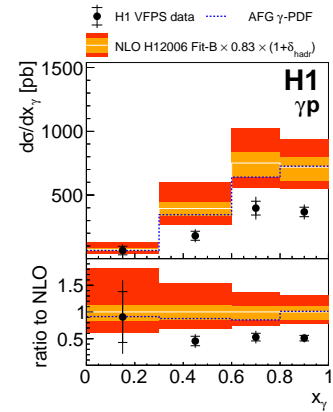,width=0.3\textwidth}
\caption{\label{fig:xgamma}Diffractive dijet photoproduction cross
  section as a function of the variable $x_\gamma$.}
\end{center}
\end{wrapfigure}
Single-differential cross sections are measured as a function of
various variables. For $\gamma p$ it is particularly interesting
to look at the variable $x_\gamma$, which is sensitive to the
longitudinal momentum fraction of the photon entering the hard
process. In a simplified view, the photon has a hadronic structure at
small $x_\gamma$ but has no structure at $x_\gamma=1$.
As can be seen in figure \ref{fig:xgamma}, the shape in $x_\gamma$
is well described by the NLO calculation, but the normalisation is
not.
This observation is consistent with earlier measurements \cite{Aktas:2007hn,Chekanov:2007rh,Aaron:2010su}.
\begin{wraptable}{r}{0.45\textwidth}
{\footnotesize
\begin{center}
\begin{tabular}{l|cc}
\hline
Systematic uncertainty & DIS & $\gamma p$ \\
\hline
VFPS detector & $3.0\%$ & $5.3\%$ \\
Hadronic energy scale & $4.4\%$ & $7.2\%$ \\
RAPGAP model & $4.3\%$ & $6.9\%$ \\
Normalisation & $6.0\%$ & $6.0\%$ \\
\hline
Total & $9.1\%$ & $12.8\%$ \\
\hline
\end{tabular}
\end{center}
}
\caption{\label{tab:syserror}Summary of systematic uncertainties.}
\end{wraptable}
Integrated over the full phase-space, the cross section for
dijet production in diffractive DIS is measured to be
$30.5\pm1.6\text{(stat)}\pm 2.8\text(syst)\,\text{pb}$, in agreement with the NLO
prediction of
$28.3^{+11.4}_{-6.4}\text{(scale)}^{+3.0}_{-4.0}\text{(DPDF)}\pm0.8\text{(hadr)}$.
In $\gamma p$, however, the cross section is measured to be 
$237\pm14\text{(stat)}\pm 31\text(syst)$ whereas the prediction is 
$430^{+172}_{-98}\text{(scale)}^{+48}_{-61}\text{(DPDF)}\pm13\text{(hadr)}$.
The systematic uncertainties, summarised in table \ref{tab:syserror},
stem in about equal part from the 
understanding of the VFPS detector, the hadronic energy scale, 
uncertainties of the RAPGAP simulation and the overall normalisation.

\begin{figure}[t]
\begin{center}
\epsfig{file=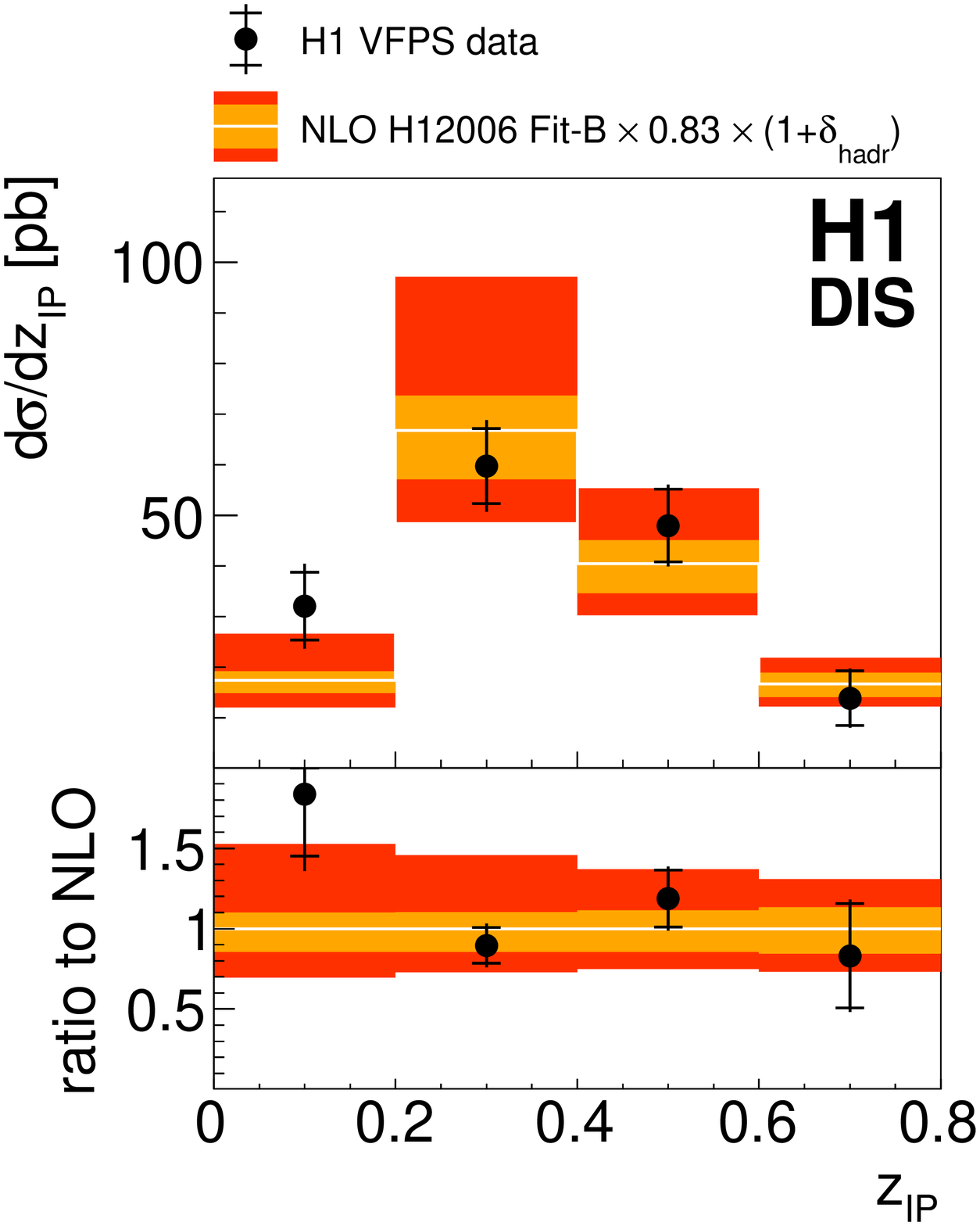,width=0.25\textwidth}%
\epsfig{file=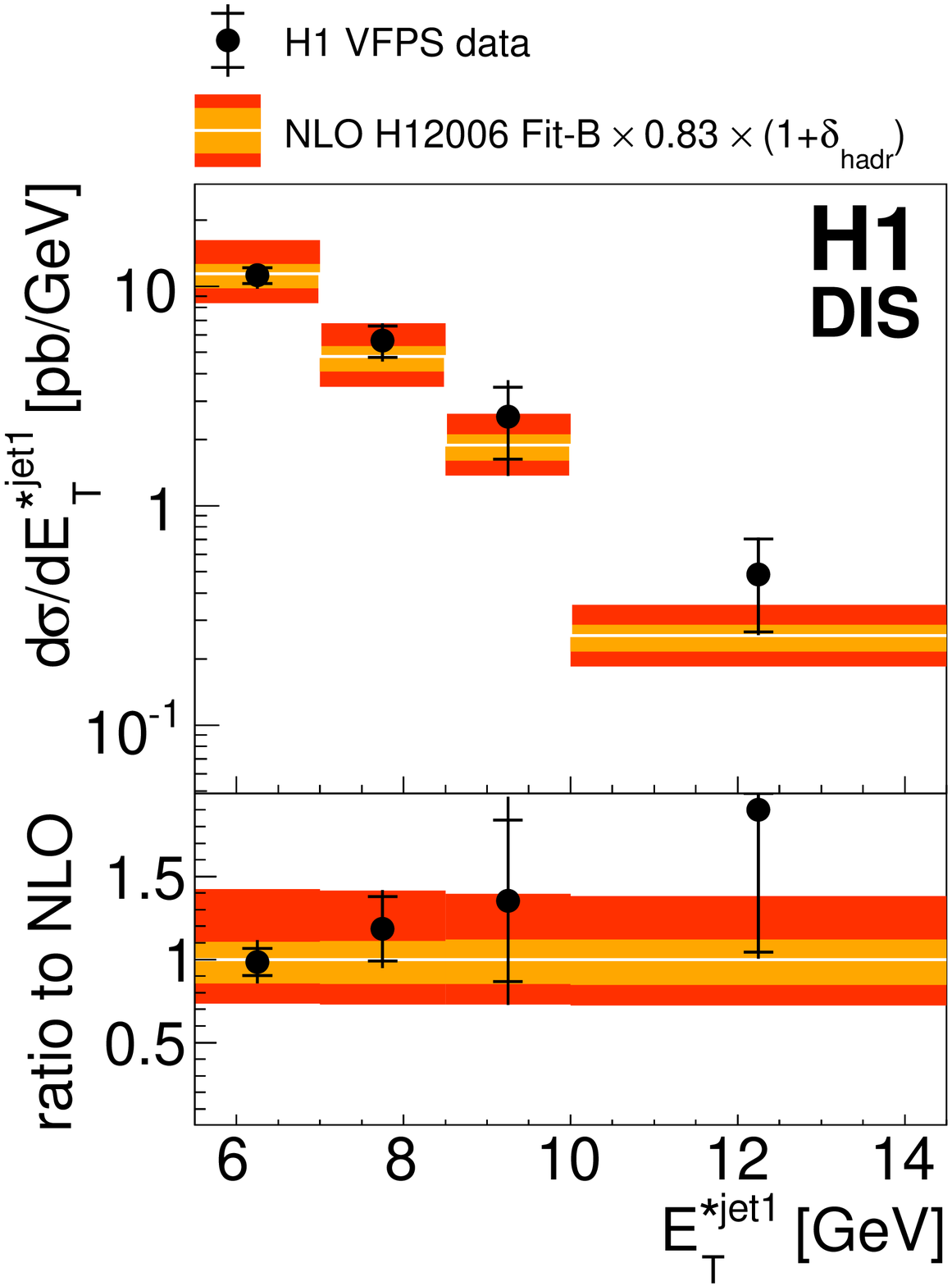,width=0.25\textwidth}%
\epsfig{file=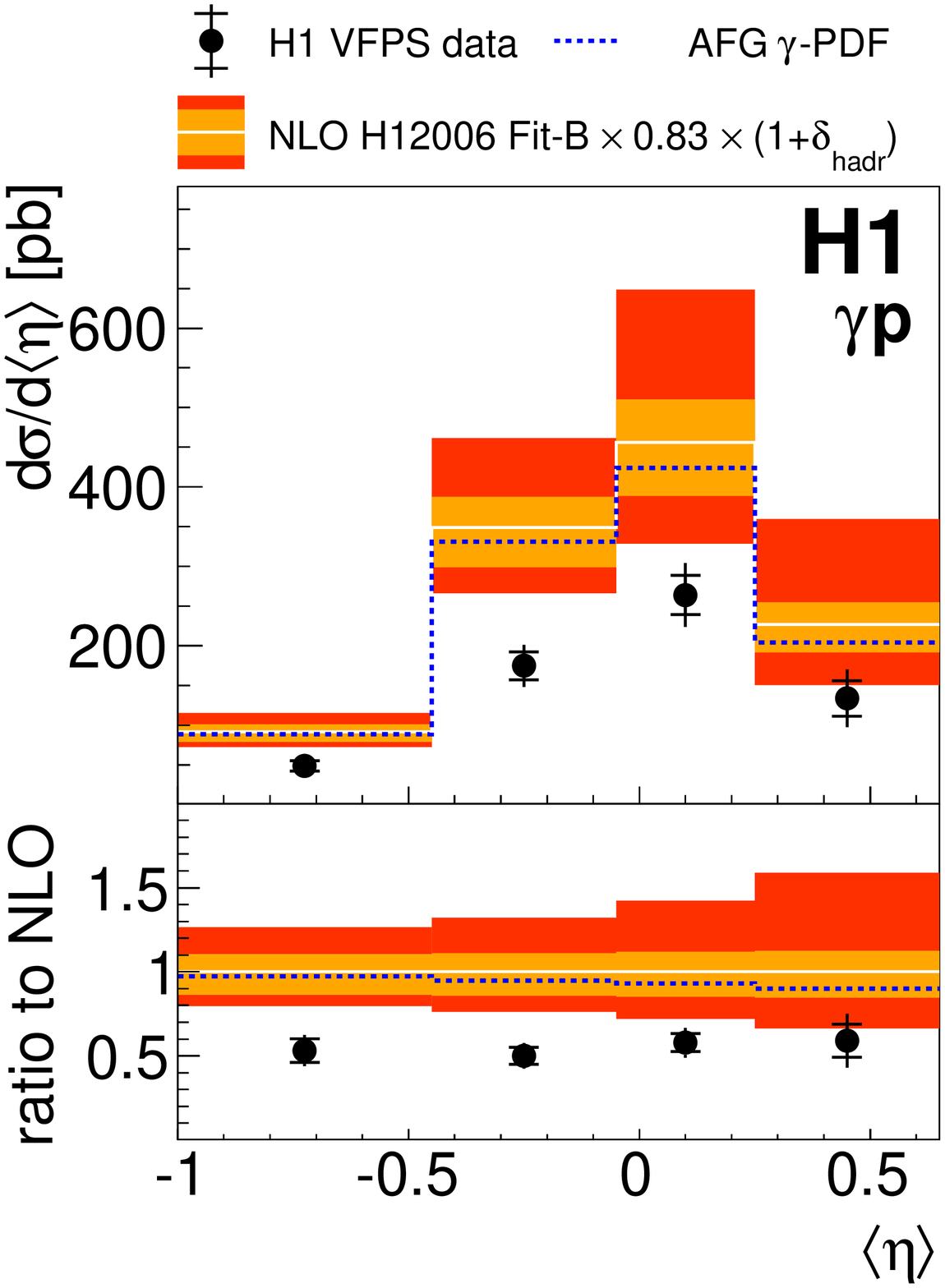,width=0.25\textwidth}
\epsfig{file=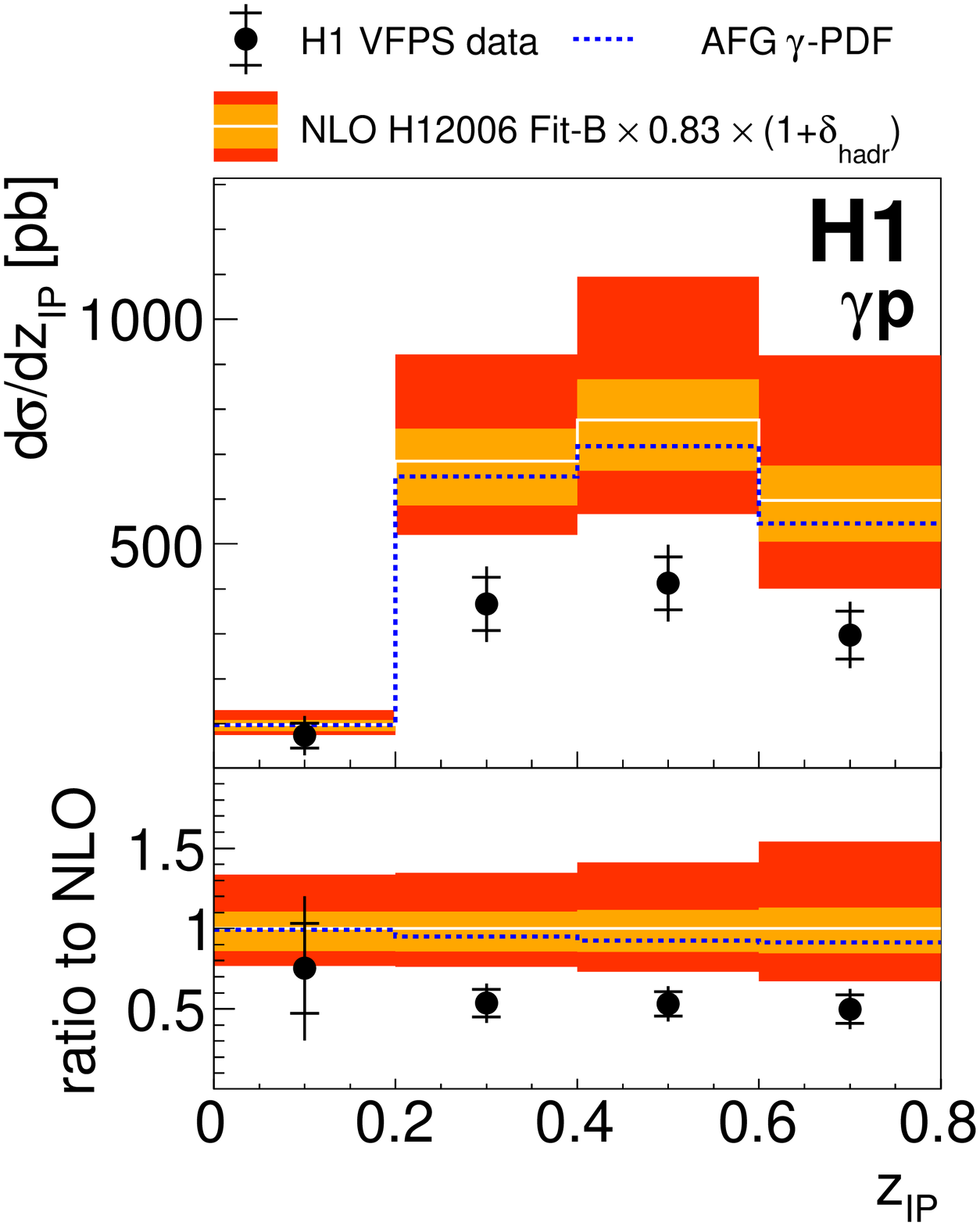,width=0.25\textwidth}%
\epsfig{file=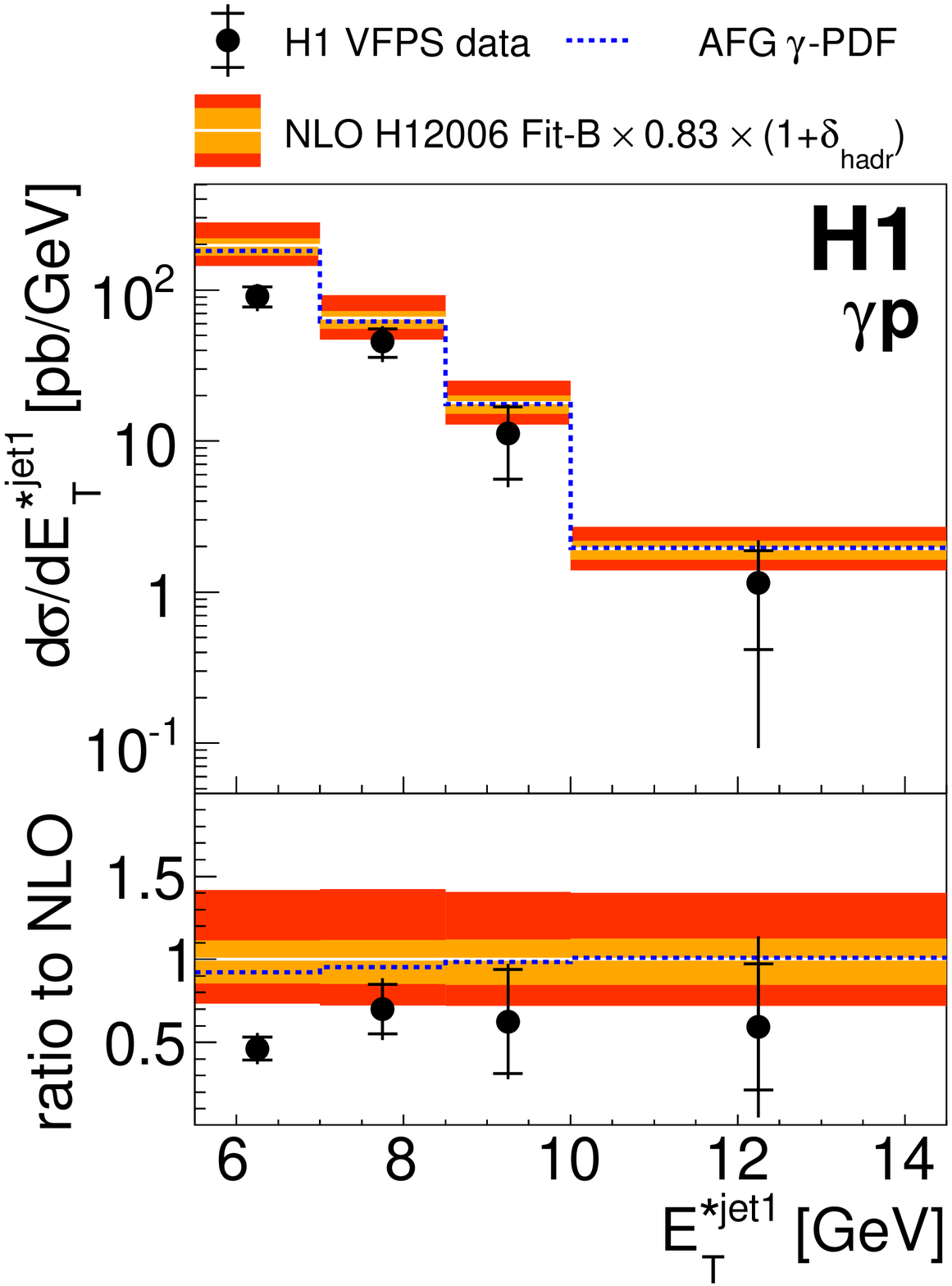,width=0.25\textwidth}%
\epsfig{file=d14-242f7d.eps,width=0.25\textwidth}
\caption{\label{fig:singlediff}Diffractive dijet production with a
  leading proton; single-differential cross sections in DIS and in
  photoproduction.
}
\end{center}
\end{figure}

\begin{wrapfigure}{r}{0.3\textwidth}
\begin{center}
\epsfig{file=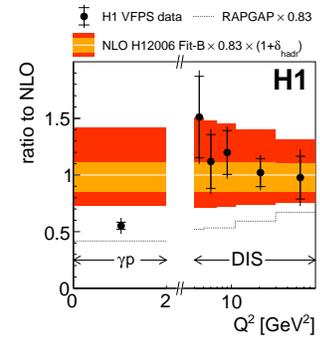,width=0.3\textwidth}
\caption{\label{fig:q2}Ratio of measured to predicted diffractive dijet cross
  section as a function of $Q^2$.}
\end{center}
\end{wrapfigure}

Figure \ref{fig:singlediff} shows the cross
sections in DIS and $\gamma p$ as a function of the variables
$z_{I\-\-P}$, $E_{T}^{\star jet1}$ and the difference in rapidity, $\vert\Delta\eta\vert$.
 In general, all single-differential cross sections are well described
 in DIS. In $\gamma p$, the shapes are well described but the overall
normalisation is off by approximately a factor of two.
This effect also can be seen in figure \ref{fig:q2}, where the
ratio of the measured to predicted cross section is depicted as a
function of $Q^2$.
In the DIS regime, the ratio is compatible with one, whereas for
photoproduction the ratio is close to one half.

\section{Cross section ratios}

In order to possibly reduce the impact of systematic uncertainties,
ratios of cross sections in $\gamma p$ to DIS and double-ratios data
to theory are also investigated.
\begin{figure}[t]
\begin{center}
\epsfig{file=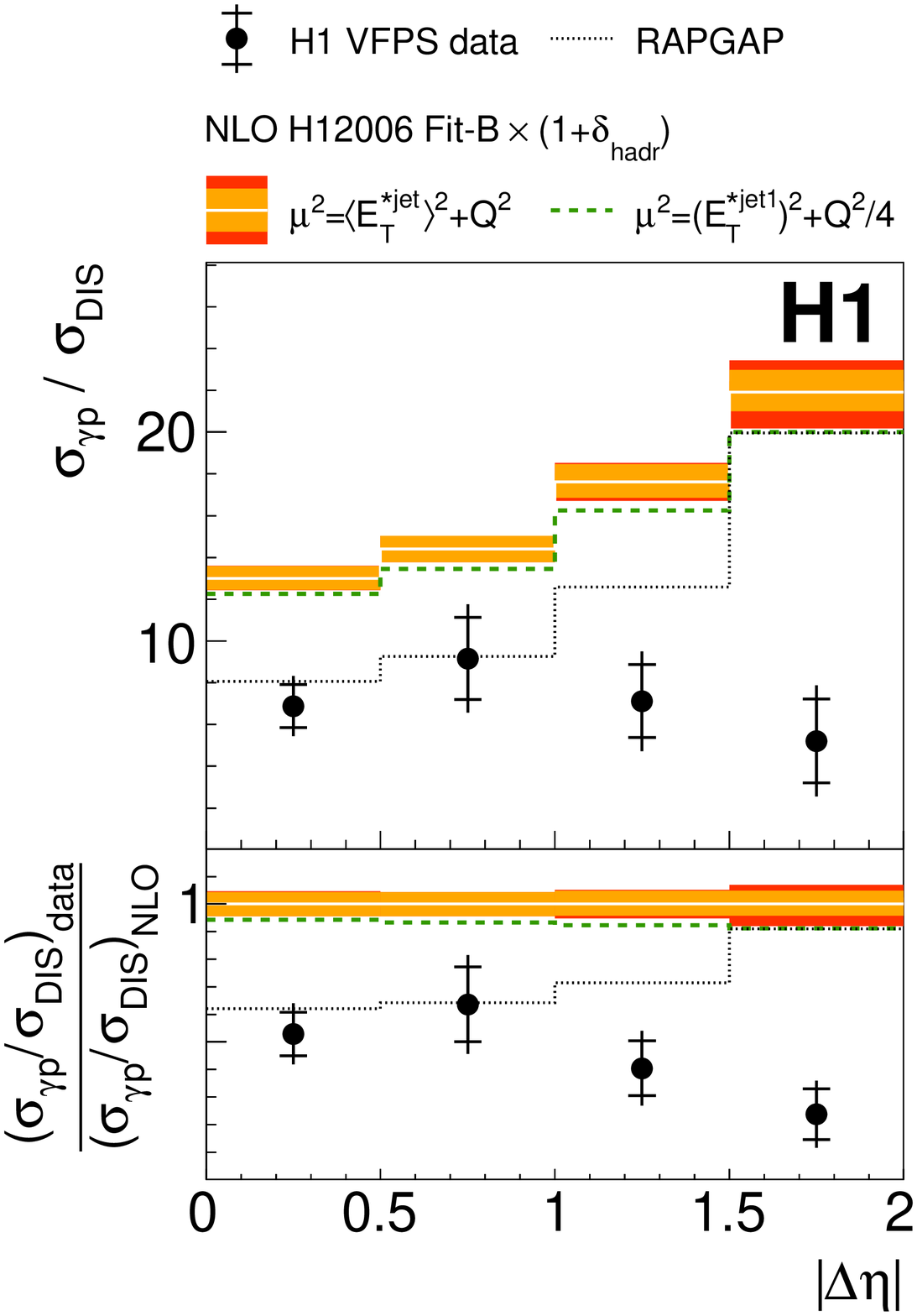,width=0.25\textwidth}%
\epsfig{file=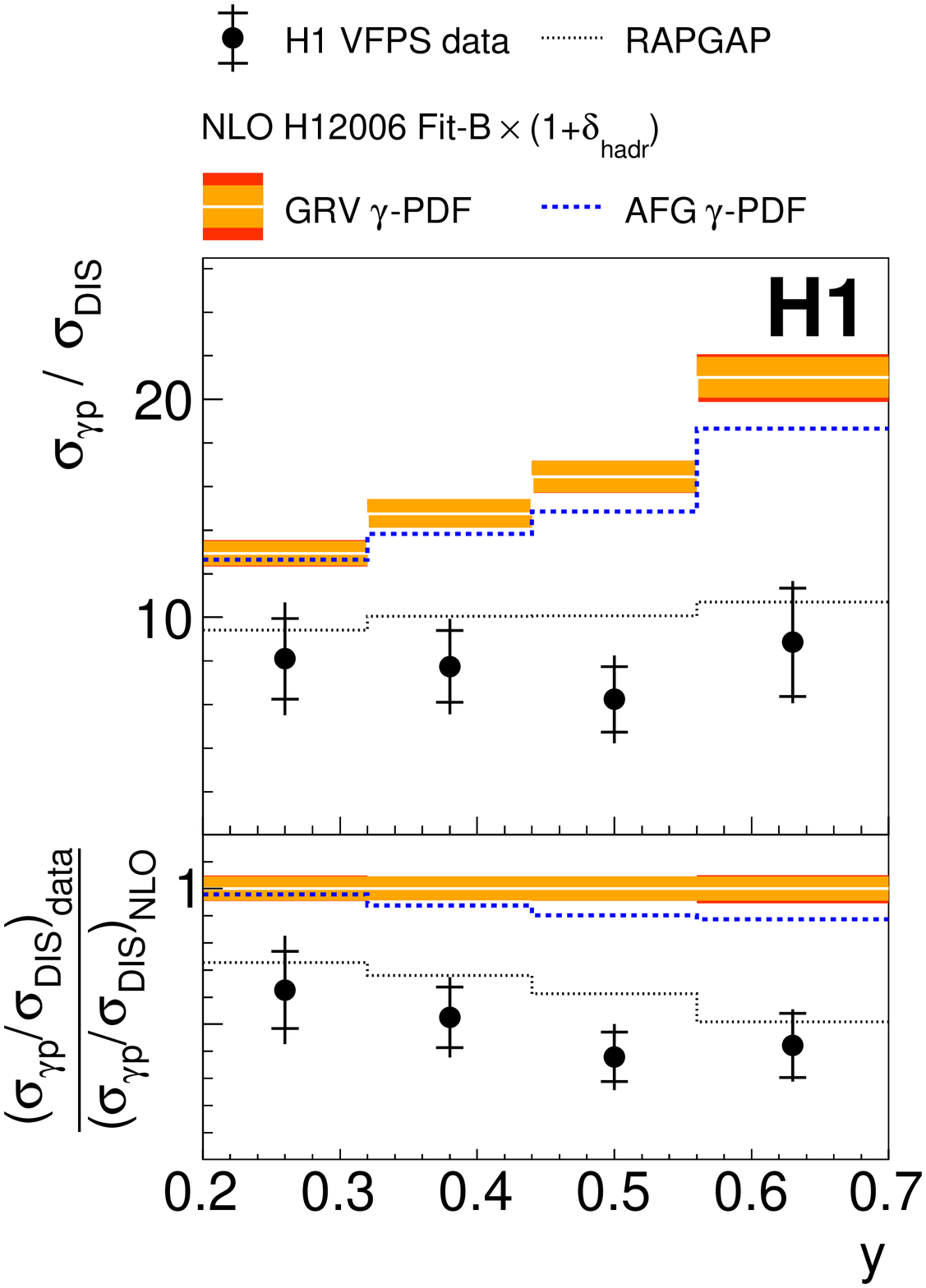,width=0.25\textwidth}%
\epsfig{file=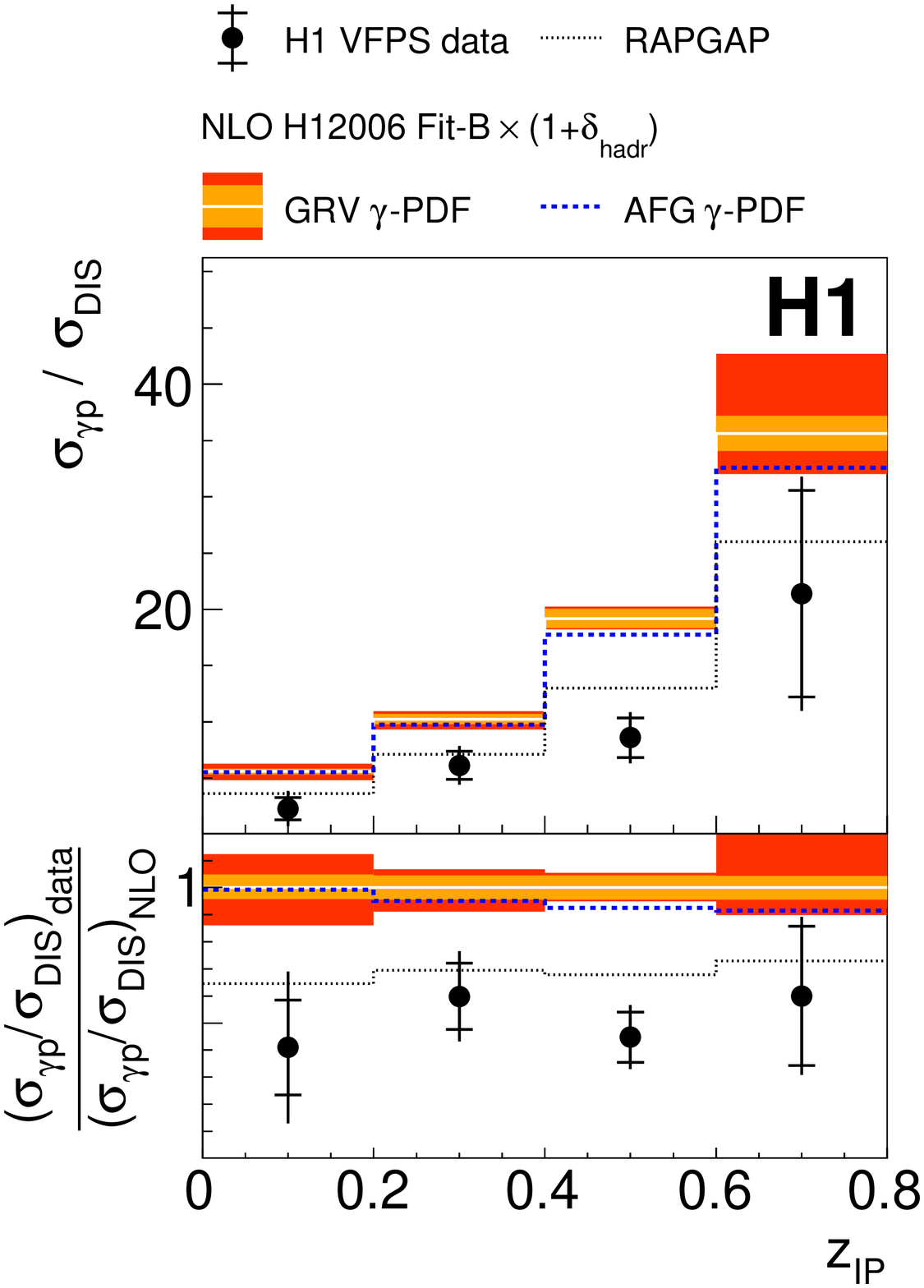,width=0.25\textwidth}%
\epsfig{file=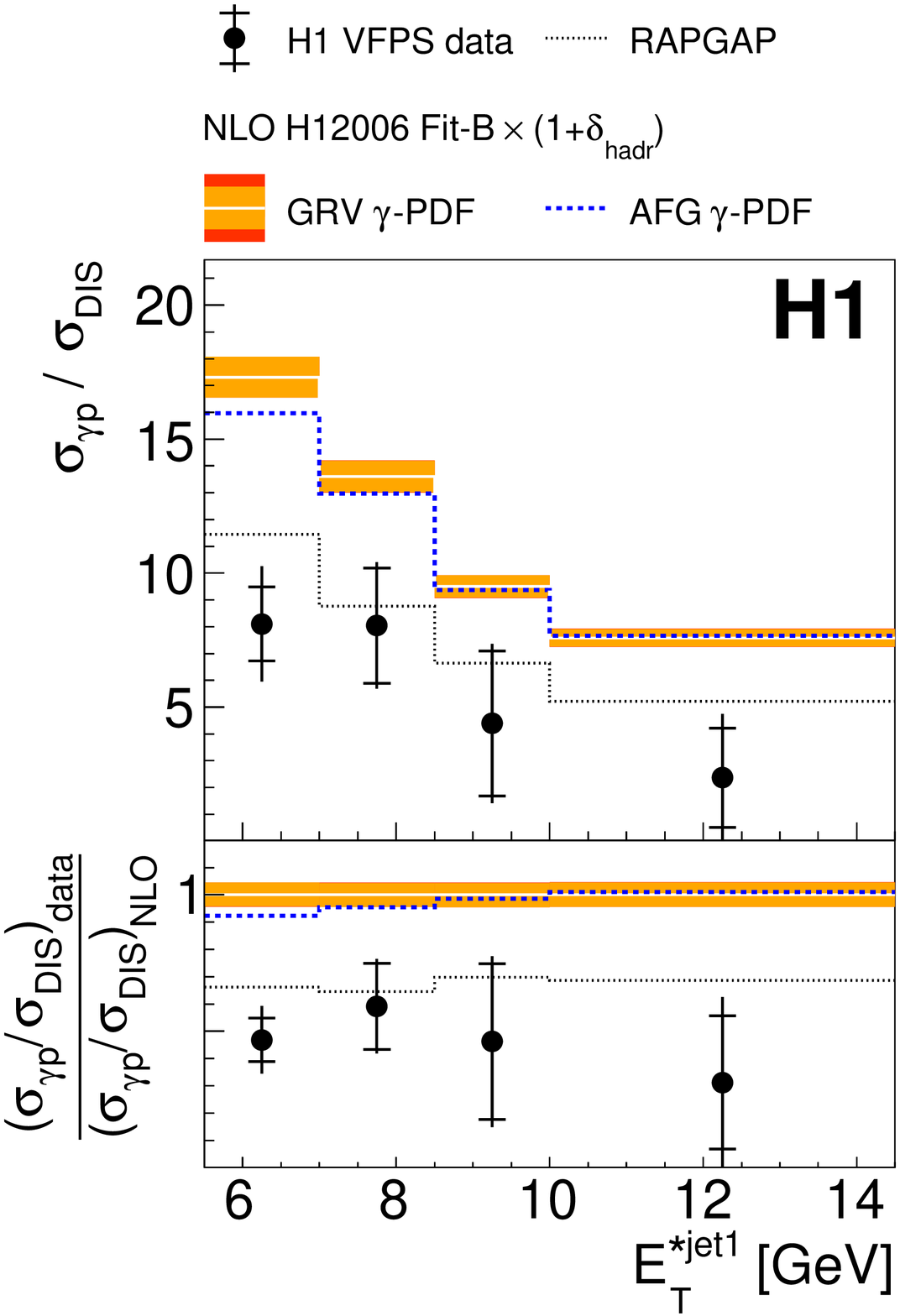,width=0.25\textwidth}
\caption{\label{fig:ratio}Ratio of $\gamma p$ to DIS diffractive dijet
  cross sections and double ratio of data to NLO prediction.}
\end{center}
\end{figure}
For the NLO 
prediction, the assumption is made that variations of the scale $\mu$
have to be applied consistently to the DIS and $\gamma p$
predictions.
This leads to cancellations of scale uncertainties in the ratio.
For the measurements, however, the systematic uncertainties do not
reduce.
The main reason are the model uncertainties, which 
are not correlated between DIS and $\gamma p$ and thus are enlarged in
the ratio.
For the integrated cross section, the double-ratio of $\gamma p$ to
DIS, data to theory is found to be $0.511\pm
0.085\text{(data)}^{+0.022}_{-0021}\text{(theory)}$.
This is consistent with an earlier H1 analysis \cite{Aktas:2007hn} of 
independent data. The present
measurement, with a tagged forward proton, is free of proton
dissociative contributions, as opposed to the earlier
measurements. Hence one may conclude that proton dissociation
certainly is not the dominant source of the suppression.

Possible shape dependencies of the cross section ratios are also
investigated. Figure \ref{fig:ratio} shows cross section ratios in the
variables $\vert \Delta\eta\vert$, $y$, $z_{I\-\-P}$ and
$E_{T}^{\star jet1}$. The double ratios are significantly different
from unity, but no significant shape dependencies are observed. The
largest deviation from a constant are observed in $\vert
\Delta\eta\vert$, however the data are still compatible with a
constant at a fit probability of $15\%$.

\section{Summary}

Diffractive dijet production with a leading proton is measured both in
the regime of deep-inelastic scattering (DIS), $Q^2>4\,\text{GeV}^2$, and in the
photoproduction regime ($\gamma p$), $Q^2<2\,\text{GeV}^2$. The data exploit for
the first time the H1 Very Forward Proton Spectrometer, which has a
large acceptance and thus small normalisation
uncertainties. Single-differential cross sections, cross-section
ratios and double ratios of $\gamma p$ to DIS are
extracted. Next-to-leading order QCD (NLO) calculations are able to describe
the measured diffractive DIS dijet cross sections, whereas they are
off in normalisation in $\gamma p$. The observed $\gamma p$ cross
sections correspond to a suppression factor of $0.511\pm
0.085\text{(data)}^{+0.022}_{-0021}\text{(theory)}$, assuming that
scale uncertainties of the NLO calculations largely cancel in the
double ratio $\gamma p$ to DIS, data to NLO. This result is in
agreement with earlier H1 measurements 
\cite{Aktas:2007hn,Aaron:2010su}.

\end{document}